 \documentclass[final,5p,times,authoryear,lefttitle]{elsarticle}

\usepackage{amssymb}
\usepackage{amsmath}

\journal{Nuclear Physics B}

\usepackage{graphicx}
\usepackage[dvipsnames]{xcolor}
\usepackage{amsmath,amssymb}
\usepackage[export]{adjustbox}
\usepackage{xcolor}

\newcommand{\Ha}{{H$\alpha$}}

\newcommand{\pab}{{Pa$\beta$}}
\newcommand{\pag}{{Pa$\gamma$}}
\newcommand{\pad}{{Pa$\delta$}}
\newcommand{\brg}{{Br$\gamma$}}

\newcommand{\Ll}{{$L_{\rm line}$}}

\newcommand{\Lacc}{{$L_{\rm acc}$}}
\newcommand{\Macc}{{$\dot{M}_{\rm acc}$}}
\newcommand{\Mout}{{$\dot{M}_{\rm out}$}}

\newcommand{\Msun}{{$M_{\odot}$}}

\newcommand{\Zsun}{{$Z_{\odot}$}}

\newcommand{\Mstar}{{$M_{\star}$}}
\newcommand{\Lstar}{{$L_{\star}$}}
\newcommand{\Rstar}{{$R_{\star}$}}
\newcommand{\Teff}{{$T_{\rm eff}$}}
\newcommand{\Av}{{$A_{\rm V}$}}
\newcommand{\Ah}{{$A_{\rm H}$}}

\newcommand{\aap}{Astronomy \& Astrophysics}
\newcommand{\apj}{Astrophysical Journal}
\newcommand{\apjl}{Astrophysical Journal letters}
\newcommand{\aj}{Astronomical Journal}
\newcommand{\apjs}{Astrophys. Journal Suppl. Series}
\newcommand{\mnras}{M.N.R.A.S.}
\newcommand{\araa}{Annual Review of Astronomy \& Astrophysics}
\newcommand{\nat}{Nature}
\DeclareRobustCommand{\ion}[2]{%
\relax\ifmmode
\ifx\testbx\f@series
{\mathbf{#1\,\mathsc{#2}}}\else
{\mathrm{#1\,\mathsc{#2}}}\fi
\else\textup{#1\,{\mdseries\textsc{#2}}}%
\fi}

\usepackage{lineno}

\begin{document}

\begin{frontmatter}




\title{SHARPing accretion and outflows in young stellar objects in star forming regions of the outer Galaxy and beyond} 


\author[1]{Juan Manuel Alcal\'a\corref{cor1} }%
\ead{juan.alcala@inaf.it}
\cortext[cor1]{Corresponding author}

\author[1]{Alessio Caratti o Garatti}
\ead{alessio.caratti@inaf.it}

\author[2]{Linda Podio}
\ead{linda.podio@inaf.it}

\author[3]{Mario Giuseppe Guarcello}
\ead{mario.guarcello@inaf.it}

\author[3]{Loredana Prisinzano}
\ead{loredana.prisinzano@inaf.it}

\author[3]{Rosaria Bonito}
\ead{rosaria.bonito@inaf.it}

\author[4]{Paolo Franzetti}
\ead{paolo.franzetti@inaf.it}

\author[1]{Fedor Getman}
\ead{fedor.getman@inaf.it}


\affiliation[1]{organization={INAF-Osservatorio Astronomico di Capodimonte},
addressline={Via Moiariello 16},
postcode={I-80131},
city={Napoli},
country={Italy}}

\affiliation[2]{organization={INAF-Osservatorio Astrofisico di Arcetri},
addressline={Largo E. Fermi 5},
postcode={I-50125},
city={Firenze},
country={Italy}}

\affiliation[3]{organization={INAF–Osservatorio Astronomico di Palermo},
addressline={Piazza del Parlamento 1},
postcode={I-90129},
city={Palermo},
country={Italy}}

\affiliation[4]{organization={INAF–IASF Milano},
addressline={via Bassini 15},
postcode={20133},
city={Milano},
country={Italy}}

\begin{abstract}
As part of the science book of SHARP, we present here the science case of star-disk interaction of low-mass (\Mstar$\leq$2\Msun) young stellar objects (YSOs), in low-metallicity (Z$<$ 0.2 \Zsun) star forming regions (SFRs) and supermassive star clusters, using the SHARP instrument mounted on the ESO-ELT. 
Extreme adaptive optics (AOs),  with a spatial resolution a factor $\sim$3 better than JWST, as well as sensitive multiplexing capabilities, uniquely offered by SHARP, are essential to efficiently survey 
the whole area of low-Z SFRs and massive clusters in the outer Milky Way (MW) Galaxy and in the Magellanic Clouds (MCs).
Using the SHARP exposure time calculator (ETC) we demonstrate that SHARP can achieve the required signal-to-noise, both for the continuum and emission lines, to investigate accretion and outflows in YSOs in distant (d$>$5\,kpc) SFRs, including those relatively embedded. SHARP will be able to observe very faint YSOs ($H\sim$\,24\,mag), allowing us extending studies to very low-mass YSOs in distant SFRs. The performance of SHARP in terms of sensitivity and spatial resolution in the NIR will provide significant insights into the evolution of protoplanetary disks in low-metallicity and massive environments: studies of accretion, jets/winds and photo-evaporation processes, down to the very low-mass ($\sim$0.2\,\Msun) regime in the MCs, and down to substellar YSOs in SFRs of the outer MW Galaxy (d\,$\lesssim$\,10\,kpc), will be possible. SHARP will also be able to observe  jets/outflows in targets that are several magnitudes fainter than those reachable with current instruments, and will facilitate studies in low metallicity environments of wide binaries and multiple systems, with separations of $\sim$1600\,au, at a distance $\sim$50\,kpc scale, and of $\sim$150\,au, in regions of the outer MW Galaxy (d $\sim$10\,kpc).

\end{abstract}




\end{frontmatter}



\section{Introduction}
\label{Intro}


Much of what we know today about low-mass star formation comes from IR imaging surveys of Young Stellar Objects (YSOs) in nearby star forming regions \citep[SFRs,  see for instance][and references therein]{evans09, dunham15}, while spectroscopic surveys of YSOs in nearby (d < 500\,pc) SFRs, have investigated
the interplay between accretion, jets and disk structure at different masses and evolutionary stages  (\citealt{alcala17, frasca17, giannini15, manara17, nisini18, manara21, pittman22,vandishoeck25}). Due to the low sensitivity and the limited spatial resolution currently available, many other aspects are still unexplored. One example is the study of low-mass YSOs in distant ($\sim$ kpc scale) SFRs, where the low-metallicity and stellar environment effects may have an important impact on the accretion/ejection process. This research is in its infancy at best (\citealt{demarchi24}, and references therein). This also includes very massive young clusters with a mass greater than some 10$^4\,$M$_\odot$, named supermassive star clusters (SSCs), hosting rich ensembles of massive stars that create environments dominated by energetic radiation, which affects the dispersal and evolution of protoplanetary disks. \citep[e.g.][]{2025OJAp....8E..54A}.

A debated point is that young ($\sim$1-3\,Myr) stellar objects in Galactic SFRs at low-Z ($<$ 0.2 \Zsun) disperse their circumstellar disks more rapidly than their solar-Z analogs (\citealt{yasui21}). This suggests that low-Z may induce higher mass accretion rates (\Macc) and/or stronger winds or very efficient disk photo-evaporation, both processes yielding a faster disk dispersal. Disk dispersal timescales are also well known to depend on the intensity of the UV radiation emitted by nearby massive stars and incident on protoplanetary disks, as a consequence of the photoevaporation process \citep[e.g.,][and references therein]{2022EPJP..137.1132W}. This process, driven by UV radiation incident on the disks, creates a flow of gas away from the disk, thereby reducing the reservoir of material and resulting in faster dispersal, which in the most extreme cases can occur on timescales as short as 1$\,$Myr.  
As a consequence, most YSOs forming in a low-Z or massive environment should experience disk dispersal earlier (in $\leq$1 \,Myr) than those forming in \Zsun\ environments or low-mass clusters, where disk dispersal may last up to 6-8\,Myr (\citealt{yasui23}). This is in contrast with recent results from JWST studies (\citealt{demarchi24}) that hint to much longer disk lifetimes in low-Z SFRs. Thus, the environment plays a pivotal role in circumstellar disk evolution, hence on the planet formation timescale and on the (debated) dependence of exoplanet frequency and chemical composition in metal-rich stars. 

In this contribution, part of the SHARP science book (see Saracco in this issue), we present the SHARP science case on the characterization of accretion and jets/outflows in YSOs in SFRs in the most massive Galactic young clusters and in the outer Milky Way (MW) Galaxy and the Magellanic Clouds (MCs).
Methodologies to investigate the star-disk interaction, as well as previous work are presented in Section~\ref{methods_previouswork}. In Section~\ref{proposal} our science case of observing low-mass YSOs in supermassive clusters, outer MW Galaxy and in the MCs  with SHARP is presented, justifying the need for sensitive and high-spatial resolution observations, and reporting the SHARP performance for studies of low-mass YSOs in distant SFRs; the goals of the science case are also presented in this section, while our conclusions are presented in Section~\ref{conclusion}.

\section{Methodologies and previous work}
\label{methods_previouswork}

A key step to determine the star-disk interaction is to first characterize its YSOs in terms of their physical properties. Therefore, we focus here on these matters by comprehensively summarizing  methodologies to determine fundamental stellar parameters, as well as mass accretion and ejection, which are key processes for star and planet formation. We also summarize some previous studies of YSOs in low-Z and supermassive SFRs.

\subsection{Stellar properties}
The near-infrared (NIR) ($\sim0.9-2.5~\mu m$) regime at medium (few thousands) spectral resolution already grants to accurately derive fundamental stellar parameters, such as the effective temperature (\Teff) of YSOs in distant, highly extinguished and low-metallicity SFRs. 

A primary challenge in deriving \Teff\ in actively accreting low-mass YSOs, is the presence of a complex, multi-component contribution. In fact, in addition to the stellar photospheric contribution, the YSOs emission also includes components from magnetically driven accretion and ejection processes, and from cool star spots (e.\,g. \citealt{gangi22}, and references therein). 
Magnetic fields in low-mass YSOs can modify stellar interiors, producing radius inflation, inhibiting convection, altering effective temperature, and causing departures from standard pre-main-sequence evolutionary tracks \citep[see][]{some17, jeff21, fran22, Cao25}. 
Accounting for this composite stellar emission is essential, as the conventional method of subtracting a single - temperature spectral 
template is often insufficient (see \citealt{gangi22}). The NIR spectral range is particularly informative because  the relative 
contribution of starspot emission is maximized with respect to the unspotted stellar photosphere (see \citealt{pere25}).

A determination of the $T_{\mathrm{eff}}$, through the modeling of the NIR photospheric spectrum, is essential. Several temperature-sensitive photospheric lines like the  \ion{Na}{i} doublet 
and the \ion{Ca}{i} triplet 
in the K-band, can be fitted using model atmospheres \citep[e.g.][]{doppmann2003} and methodologies well tested in YSOs in nearby SFRs  
\citep[e.g.][and references therein]{frasca17}. Also, atomic line depth ratios from \ion{Fe}{i}, \ion{Al}{i} and \ion{OH}{} in the H-band, can be used to estimate $T_{\mathrm{eff}}$ using methodologies similar to those in the literature \citep[e.g.][]{biazzo2007}.
Note, however, that commonly used photospheric solar - metallicity tracers may be intrinsically weaker in low-metallicity YSOs. Thus, the line fitting and measurements of line depth ratios may become more challenging than for YSOs in nearby SFRs, possibly requiring development of {\it ad-hoc} procedures.

Once derived, $T_{\mathrm{eff}}$  allows one to estimate other fundamental stellar parameters such as luminosity \Lstar, radius \Rstar, mass 
\Mstar\ and age via comparison with Pre-Main Sequence (PMS) evolutionary tracks of the appropriate metallicity. Another crucial step is comparing stars on the HR diagram alongside with set of isochrones and tracks from evolutionary models for spotted stars 
(\citealt{some20,fran22}), to infer new spot-corrected age and mass. Whenever surface gravity determinations by photospheric line fitting will be possible, will allow one to make an independent determination of \Mstar.

\subsection{Mass accretion rate (\Macc)}
\label{Macc_rates}
The most efficient way to characterize YSOs and study the star-disk interaction processes is via multi-wavelength observations, in particular of CTTS. These are young (a few 10$^6$ yr), very low- to solar-mass stars that are actively accreting mass from their surrounding planet-forming disks. 

Within the magnetospheric-accretion model (\citealt{hartmann16}, and references therein) the stellar magnetic field truncates the inner disk of CTTS at a few stellar radii (\citealt{DonatiLandstreet09, johns-krull13}). The disk gas flows from this location onto the star along the magnetic field lines, leading to an accretion shock at the stellar surface. The hot (T$\sim$10$^4$\,K) gas emits in the Balmer and Paschen continua and in several permitted lines (\citealt{hartmann16}). At the same time, magnetically driven winds carry away the angular momentum of the accreting gas, preventing the star from spinning up. The accretion shocks produce strong UV and X-ray emission (e.g. \citealt{bonito14}) that irradiates and photo-evaporates the disk. 

\begin{figure*}[t]
\centering
\includegraphics[width=0.8\textwidth ]{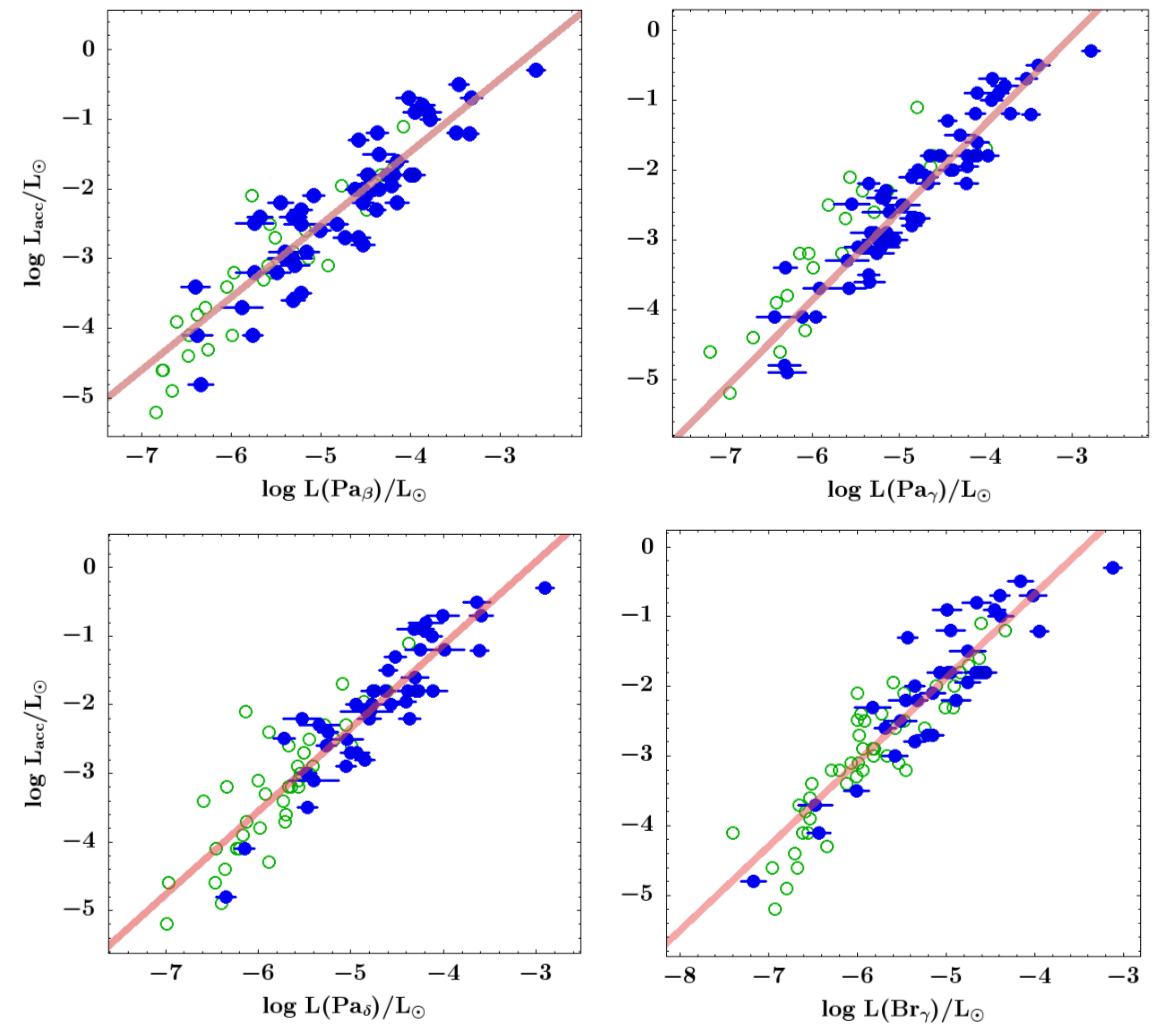}
\caption{Examples of  \Lacc -- \Ll ~relationships in the NIR for the \pab\ (upper left), \pag\ (upper right), \pad\ (lower left) and \brg\ (lower right) lines. In all panels line detections are represented with the blue  dots, while upper limits with green empty circles. The linear fits are indicated with the reddish lines. Figure adapted from \cite{alcala17}.}
\label{Lacc-Ll-rel} 
\end{figure*}

The energy released per unit time in the accretion shock, i.e. accretion luminosity (\Lacc), can be measured from the excess flux in the continuum and the lines relative to those of non-accreting template spectra. Such measurements are best suited at ultraviolet (UV) 
wavelengths ($\lambda$ $<$ 4000\,\AA) from the Balmer continuum excess emission and the Balmer jump  
(see  \citealt{HH08, ingleby13, alcala14, alcala17, manara17, pittman22} and references therein).  
These studies have shown  that \Lacc\ is correlated with the line luminosity, \Ll, of \ion{H}{i}, \ion{He}{i} and \ion{Ca}{ii} permitted lines (e.g. \citealt{muzerolle98, calvet04, HH08, rigliaco12, alcala14, alcala17,fiorellino25} and references therein). These works have provided \Lacc--\Ll\ empirical correlations, simultaneously and homogeneously derived from the UV to the near-infrared (NIR). The emission lines are accretion tracers hence, are key diagnostics for estimating \Lacc\ via the correlations mentioned above when flux-calibrated spectra below $\lambda$$\sim$3700\,\AA\ are not available. Examples of these correlations for several lines in the NIR are shown in Figure.~\ref{Lacc-Ll-rel}. The latter are the key tools to determine \Lacc\ from the SHARP observations. 

The accretion luminosity can be converted into mass accretion rate, \Macc, using the following equation (see \citealt{gullbring98, hartmann98})

\begin{equation}
 \label{Macc}
 \dot{M}_{acc} = \left( 1 - \frac{R_{\star}}{R_{in}} \right)^{-1}  \  \frac{L_{acc} \ R_{\star}}{G \ M_{\star}}
 \approx 1.25  \ \frac{L_{acc} \ R_{\star}}{G \ M_{\star}}
\end{equation}

\noindent
provided the stellar mass \Mstar\ and radius \Rstar, derived also from spectroscopy, are known.

\subsection{Mass ejection rate (\Mout)}
\label{Mout_rates}

The outflow activity in YSOs is an ubiquitous phenomenon and is strictly related to accretion at all mass regimes. Jets and outflows have been observed across the entire range of stellar masses, from brown dwarfs (e.g. \citealt{whelan2005}) to high-mass YSOs (e.\,g. \citealt{caratti2015}).
The study of such outflows enables us to elucidate the relationship between mass ejection and accretion, and test the universality of the star formation mechanism as a function of central YSO mass.

Measurements of mass ejection, \Mout , at NIR wavelengths can be obtained from observations of typical jet/outflow tracers (e.\,g.  \citealt{dougados2010}). Among the most important are: the [\ion{C}{i}] doublet at 0.98\,$\mu$m, several [\ion{Fe}{ii}] lines in the $J$ and $H$ bands (see e.\,g. \citealt{nisini02,davis2003})\footnote{The [\ion{Fe}{ii}] lines mostly from transitions between the first 13 fine structure levels, i.\,e. the levels connecting the terms $a^6D$, $a^4F$ and $a^4D$.}, several H$_2$ ro- vibrational transitions (from $v$=1 to 4) along the whole range (see e.\,g., \citealt{davis2001,giannini04}), \ion{He}{i} at 1.08\,$\mu$m as well as many \ion{H}{i} lines from the Paschen and Brackett series. The intensity of these lines notably varies depending on the excitation conditions of the gas, being the [\ion{Fe}{ii}] at 1.26 and 1.64\,$\mu$m, and H$_2$ at 2.12\,$\mu$m the brightest (e.\,g. \citealt{nisini02,podio2006,caratti06,caratti24}). While [\ion{Fe}{ii}] lines probe the hotter collimated jet components, moving at a few hundreds km\,s$^{-1}$, H$_2$ lines mainly probe the warmer wind, moving at a few tens of km\,s$^{-1}$, although they can also be observed along the jet at the early prototstellar stage~\citep[see, e.\,g.][]{caratti06,caratti24}. From the line luminosity and the estimated jet velocity one can derive \Mout~\citep[see, e.\,g.][]{hartigan1995,podio2006,caratti2015}. 

Numerous studies have shown that mass accretion and ejection rates are strongly correlated over orders of magnitude, for both massive and low-mass YSO, at all evolutionary stages, from protostars to pre-main sequence objects (e.g., \citealt{podio2012,ellerbroeck2013, natta14,caratti2015,nisini18}). Therefore,  the mass ejection rate can also be used as a probe of accretion (e.\,g. \citealt{bonito10}) for very embedded YSO whose accretion line tracers will be obscured by the dusty envelope.

\subsection{Spectroscopy of YSOs in distant low-Z regions}
\label{previous}

Efforts to characterize the solar-mass YSO population in distant (d $<$ 10\,kpc scale) low-Z  SFRs 
of the MW Galaxy have been done during the last decade. Relevant studies regard the population in the Sh2-284 SFR (\citealt{cusano11, kalari15,biazzo2025}) at 4.5\,kpc.
However, because of instrumental sensitivity limits, such spectroscopic studies have been limited mainly to bright YSOs with masses higher than about 1\,\Msun.  Using Chandra X-ray observations \cite{guarcello21} concluded that the disk evolution in the YSO members of the Dolize\,25, a cluster in Sh2-284,  is affected mainly by the low-metallicity rather than by external photo-evaporation or dynamical interactions.
ALMA and JWST observations  of  Sh2-284 (\citealt{cheng25, jadhav25}) have shown the presence of CO and H$_2$ outflows in several YSOs, indicating important star formation activity, while further studies investigate the dependence of the Initial Mass Function 
(IMF) with metallicity (\citealt{andersen25}). 
The IMF may indeed be sensitive to the environmental conditions. Compared to our detailed knowledge of the galactic stellar and sub-stellar mass function \citep[see e.g.][for a review]{bastian10}, very little is known about the low-mass (near or below 1\,\Msun) content in low-metallicity (low-Z) environments (\citealt{zinnkann24}) and in supermassive star clusters \citep{2010ARAA..48..431P}. 

Recent JWST spectroscopic observations of the NGC\,3603 cluster (\citealt{rogers25}) at a distance of 7.2\,kpc  have characterized 42 YSOs, deriving masses in the range 0.5 to 7\,M$_\odot$, with only 5 being sub-solar mass. The metallicity of the region is found to be however Solar (see \citealt{rogers25} and references therein).

Pushing to much more distant regions, \citet{demarchi24} performed JWST spectroscopy of a few accreting YSOs in the NGC 346, a SFR in the Small Magellanic Cloud, but again the sample is limited to the mass range 0.9--1.8\,\Msun.

Several of the aforementioned investigations have extensively used the \Lacc--\Ll\ relationships  derived from X-Shooter (see Section~\ref{Macc_rates}) to estimate the accretion luminosity \Lacc. In particular, the JWST studies adopted those in the NIR shown in Figure~\ref{Lacc-Ll-rel}.


Regarding outflows, 
 \cite{mcleod18, mcleod24} reported the first detection of a bipolar collimated jet, HH\,1177, in the \Ha\ line, driven by a 8\,M$_{\odot}$ star in the LMC.
 The massive YSO is also surrounded by a keplerian disk, which makes this system similar to disk-jet systems observed in the Milky Way (e.g. \citealt{floresrivera23, bacciotti25}). At difference with accreting YSO in the MW, however, the YSO driving HH-1177 is not deeply embedded in its natal envelope, which makes it visible at optical wavelengths.  As discussed by \cite{mcleod24} this may be due to formation in a low-metallicity hence dust poor environment, thus opening the way to the detection of several disk-jet systems in the near-infrared with SHARP, even in embedded protostellar systems.

Finally, H$_2$ rovibrational transitions have been detected in the JWST NIRSpec spectra of PMS objects in the LMC, which are indicative of mass ejection driven by these sources (\citealt{demarchi24}).

\subsection{Studies of YSOs in supermassive star forming regions}
\label{previous2}

To date, the impact of photoevaporation on disk evolution and dispersal in massive star clusters has been investigated with two approaches: spectroscopic for nearby regions and purely photometric for more distant and extinguished regions. \

Thanks to their proximity, in nearby star-forming regions hosting both young disk-bearing stars and massive members — such as the Orion Molecular Cloud, which lies at about 400$\,$pc — it has been possible to perform detailed spectroscopic analyses of evaporating disks and their surrounding envelopes shaped by incident UV radiation \citep[proplyds, e.g.][]{2024AA...687A..93A}. These studies have made a major contribution to our understanding of the impact of the environment on the physical properties of disks, such as mass, and on physical processes such as accretion and outflow. However, the picture regarding how external feedback also affects the chemistry of irradiated disks and their dust content remains much more uncertain \citep[e.g.,][]{2023ApJ...958L..30R}. \

However, the Solar neighborhood only hosts low-mass star-forming regions with moderate local UV fluxes and sparse populations of stars with disks. To understand how external feedback globally impacts the evolution of the population of stars with disks in massive clusters, and to extend our knowledge to epochs when our Galaxy formed stars more actively, it is essential to study the disk population in supermassive star clusters. These clusters lie at distances of several kpc, are typically obscured by visual extinction exceeding 10 magnitudes, and suffer from severe source crowding, requiring the use of telescopes with high sensitivity, spatial and spectral resolution, and operating in the infrared.

\section{Proposed observations with SHARP}
\label{proposal}

\subsection{Need for spatial resolution and sensitivity}
Because of limiting sensitivity, previous spectroscopy of YSOs in supermassive clusters and in the SFRs of the outer MW Galaxy and the MCs had the inconvenient
of a low S/N, besides the difficulties of background subtraction. Such observations have been limited so far to the brightest, most massive YSOs 
(see Section~\ref{previous}).
In addition, even in the case of sensitive JWST IR observations of very distant SFRs, the data are sometimes difficult to interpret,  because of the 
possible mutual contamination by the spectrum of several unresolved objects falling on the same entrance slit or fiber bundle.

Due to crowding and low flux limits ($\sim$ 10$^{-18}$ --10$^{-21}$ erg / s /cm$^2$)\footnote{Estimated fluxes are based on  measurements of NIR emission lines in low-mass YSOs in the Lupus SFR, which are in the range 10$^{-13}$ -- 10$^{-16}$\,erg/s/cm$^2$ (see \citealp{alcala14, alcala17}).} of NIR emission lines,  a good Strehl ratio ($>$0.6) and high sensitivity are required in order to address the star-disk interaction issues of YSOs in low-metallicity and massive distant (greater than several kpc) regions. In the following we test the observability of YSOs in the MCs, which are the most distant targets important for this science case.

Typical MW embedded clusters have a size ranging from 1\,pc to  4\,pc, similar to LMC clusters, although those in the
MW tend to be smaller than in the LMC (see \citealt{romita16}).  A typical low-mass SFR in the solar neighborhood (e.\,g. Taurus, Lupus, or the Orion cluster; d$\sim$ 140--400\,pc) would extend $\sim$\,4\,arcsec at the distance of the MCs. 
The star density of a SFR like the Trapezium cluster is on the order of 2--4$\times$10$^4$ stars/pc$^3$ (see \citealt{bate03}, and references therein), while more recent studies give a surface density of $\sim$7.2$\times$10$^3$ stars/pc$^2$ (\citealt{rivilla13}). Such cluster surface density would be $\approx$\,420\,stars/arcsec$^2$ at the distance of the MCs. For comparison, each unit of SHARP mIFU-Vesper will cover an area of 2.5\,arcsec$^2$ hence, each of them may in principle observe, in one shot,  more than
10$^3$ stars of a cluster similar to the Trapezium, but in the MCs.  

\begin{figure*}[t]
\centering
\includegraphics[scale=0.4]{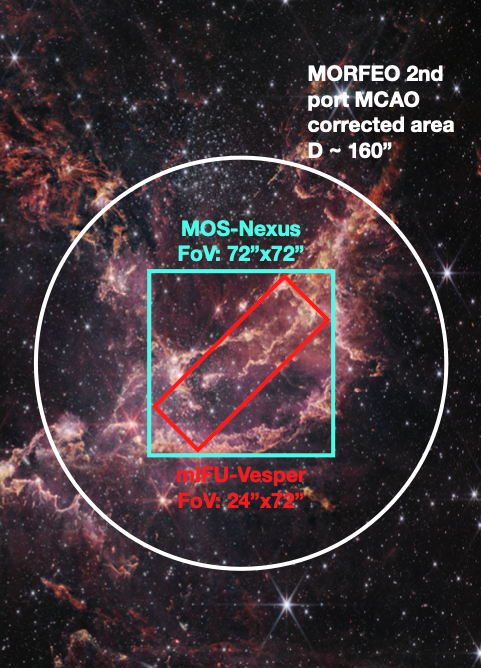}
\caption{NGC 346, a SFR in the Small Magellanic Cloud, observed with NIRCam onboard JWST, with overlapped the MORFEO MCAO corrected area (D$\sim 160"$, in white), the FoV of the MOS-Nexus ($72"\times72"$, in cyan), and the FoV of the mIFU-Vesper ($24"\times72"$, in red). Credits: NASA, ESA, CSA, Olivia Jones (UK ATC), Guido De Marchi (ESTEC), Margaret Meixner (USRA); Image Processing: Alyssa Pagan (STScI), Nolan Habel (USRA), Laura Lenkić (USRA), Laurie Chu (NASA Ames).}
\label{fig:ngc346}
\end{figure*}

Figure \ref{fig:ngc346} shows the potential of studying this kind of regions with SHARP: the field-of-view (FoV) of MOS-Nexus ($72"\times72"$) and mIFU-Vesper ($24"\times72"$) are perfectly suited to simultaneously obtain the spectra of several YSOs in NGC 346, a SFR in the Small Magellanic Cloud, recently observed with JWST \citep[e.g., ][]{demarchi24}. 

The resolving power between 2000--17000 of SHARP on ELT will allow us to spectrally resolve emission lines in YSOs for kinematical studies. In addition, with a spatial resolution $\sim$3 times better than JWST, and high sensitivity (see Section~\ref{performance}) SHARP on ELT is the ideal instrument for the characterization of multiple YSOs systems in the outer MW Galaxy and in the MCs. At the distance of 50\,kpc SHARP will be able to resolve binaries with a separation of $\sim$1550\,au, i.e. wide binary systems, while in regions like Sh2-284 or the almost completely unexplored supermassive star clusters RSGC1-3, about ten times closer (see Section~\ref{previous}) SHARP will allow us to enter into the realm of low-Z, low-mass close binaries and shed light on how the dense stellar environment in supermassive clusters affects binarity in low-mass stars.

Regarding the jets and outflows, at a distance of 50\,kpc and with a slit length of 2.5" ($\sim$0.6\,pc), SHARP-Nexus will be able to partially or fully cover the extend of jets and outflows from young stars and protostars, as the extent of protostellar jets varies from a few tenths to a few parsecs, depending on the age of the source. Given the pixel scale, single knots along the jets will not be spatially resolved, but the many different lines tracing winds, outflows and jets will be spectrally resolved. Indeed, Nexus spectral range (0.95-2.45\,$\mu$m) encompasses several bright and important emission lines from protostellar jets and outflows, which can be used as a diagnostic of the gas physical, kinematical and dynamical properties (see Section~\ref{Macc_rates}).  

Typical line fluxes in nearby (300-400\,pc) jets (with \Av $\sim$\ 10\,mag) range from 10$^{-13}$ to 10$^{-16}$\,erg\,s$^{-1}$\,cm$^{-2}$ (see e.\,g. \citealt{nisini02,giannini04,caratti06,podio2006}). This would translate to few 10$^{-18}$ down to 10$^{-21}$\,erg\,s$^{-1}$\,cm$^{-2}$ at the distance of MCs, but, considering the low-metallicity, the line flux of iron and other atomic species may be further reduced by about one third. These flux levels can be well detected with SHARP (see subsection~\ref{etc_results}.) In conclusion, SHARP-Nexus will allow us, for the first time, to detect and study low-mass protostellar jets and outflows in regions of low-metallicity, permitting to study their physical, kinematic, and dynamical properties and to compare them with those from protostars with solar metallicity, possibly opening a link with star formation studies in high-redshift galaxies.


\begin{figure*}[t]
\centering
\includegraphics[width=1.0\textwidth ]{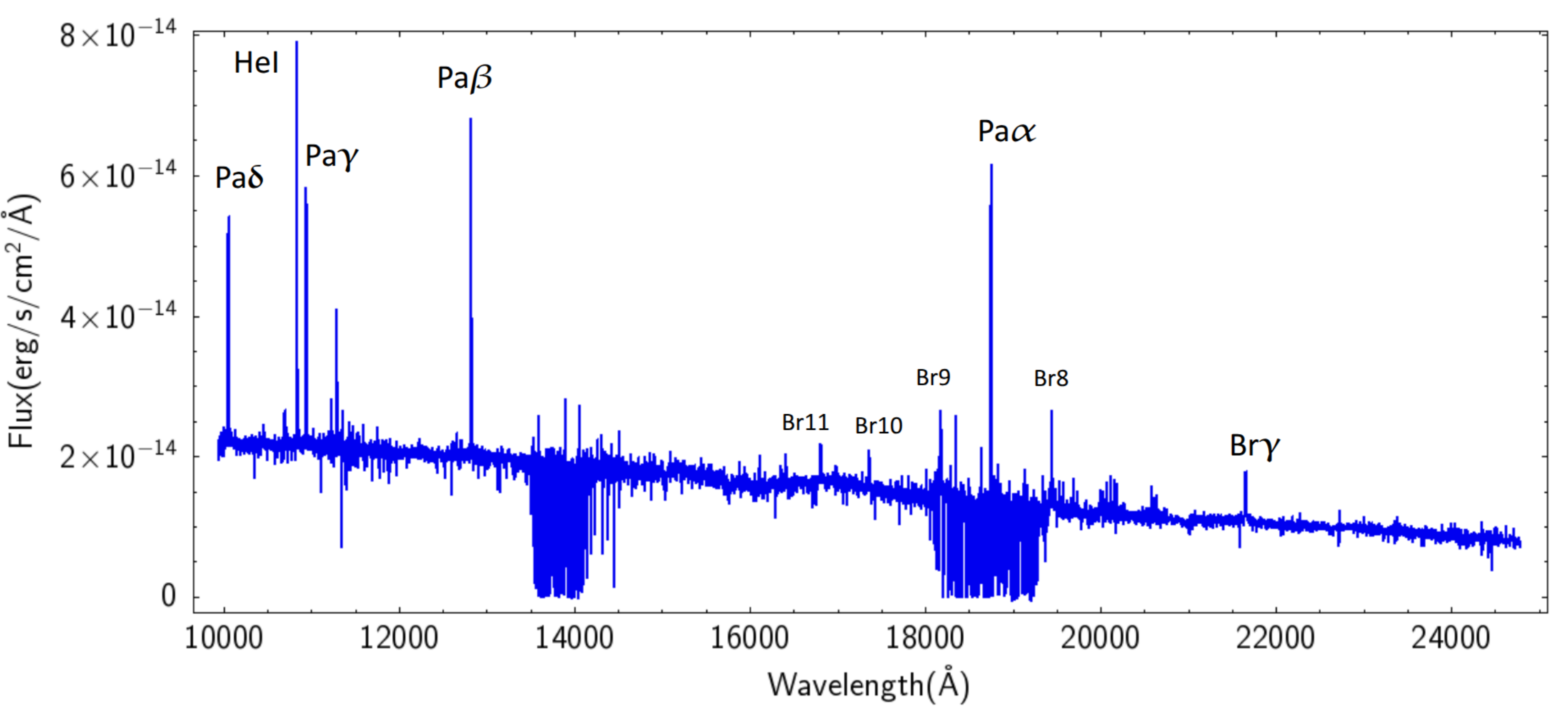}
\caption{X-Shooter spectrum of the CTTS Sz\,88\,A in the NIR, used as template for the SHARP ETC. Several of the most prominent accretion tracers are marked.}
\label{template_Sz88A}
\end{figure*}

\begin{figure*}[t]
\centering
\includegraphics[width=0.7\textwidth ]{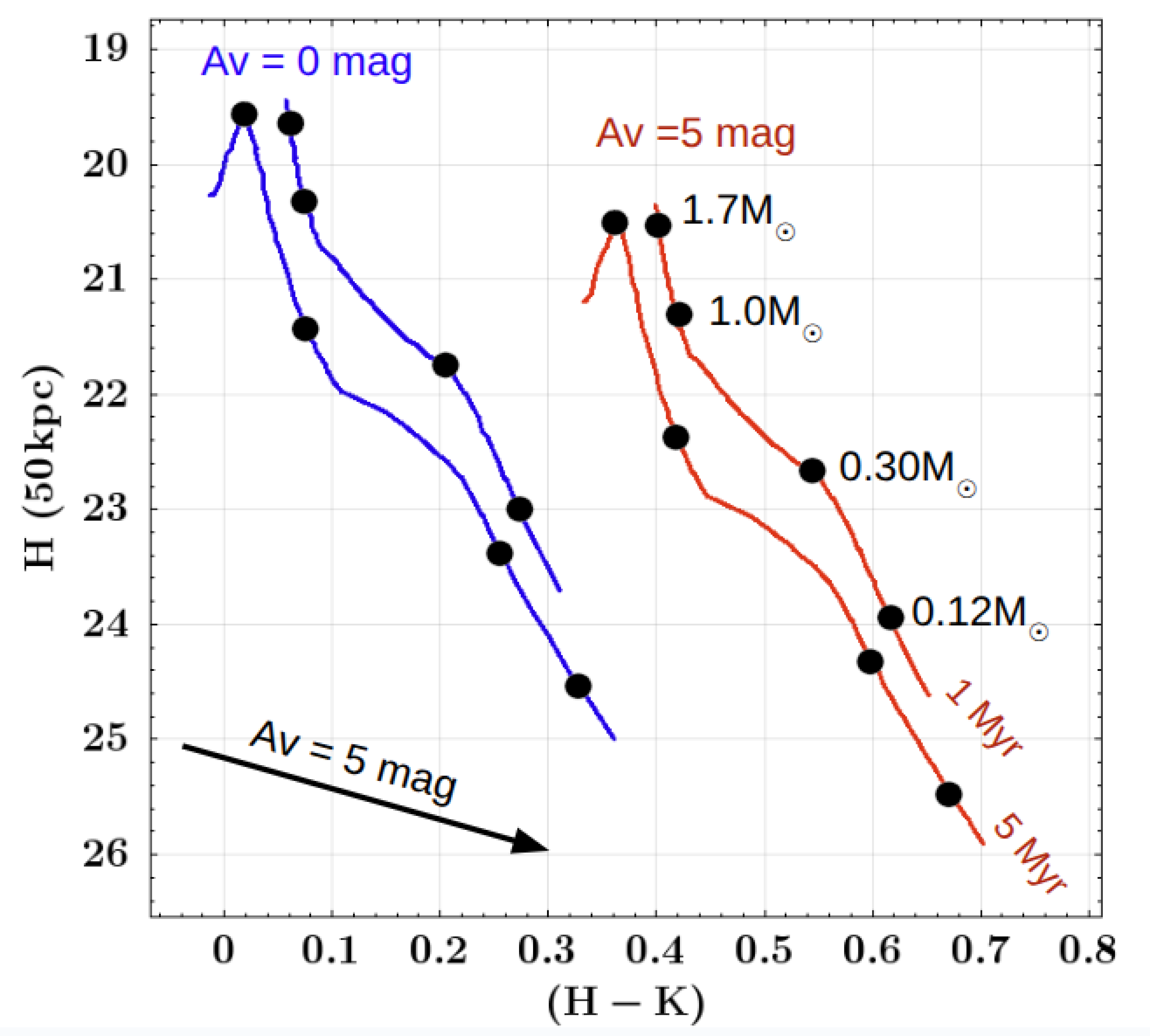}
\caption{$H$ vs. $H-K$ diagram of theoretical 1 and 5\,Myr isochrones with [M/H]$=-0.5$, and with the $H$ magnitude
shifted at a distance of 50\,kpc. The isochrones, drawn from the PARSEC tracks V 1.2S (\citealt{bressan12}, see also the web site https://stev.oapd.inaf.it/PARSEC/tools.html), are shown for two values of extinction namely, \Av$=0$\,mag and \Av$=5$\,mag in blue and red, respectively. Various YSO masses are labeled on the reddened 1\,Myr isochrone and correspondingly marked
in all isochrones with the black dots. The 5\,mag reddening vector, corresponding to \Ah$\sim$1\,mag. is shown in the 
lower left. }
\label{col-mag-diag}
\end{figure*}

\subsection{SHARP performance for studies of low-mass low-Z YSOs in distant SFRs}
\label{performance}

To investigate the instrument performance for the YSOs case we used the exposure time calculator (ETC)  developed for SHARP (\citealt{saracco24})\footnote{https://sharp.brera.inaf.it/tools/}. To apply this tool to our science case, we used VLT/X-Shooter 
spectra as input templates. These spectra simultaneously cover a wide spectral range, from the UV to the NIR,  and are flux-calibrated
at the 5-10\% level (see \citealt{alcala17, manara21}). We used in particular those of the CTTS  Sz\,88\,A. An example of template 
spectrum is shown in Figure~\ref{template_Sz88A}. We fixed ETC parameters as follows: 

\begin{itemize}
\item nominal exposure time of 1\,hour (10\,NDIT $\times$ 360\,sec)
\item the air-mass at 1.5
\item slit width of 35\,mas for Nexus
\item use the multi-conjugate AO option 
\end{itemize}

For the purpose of mass accretion  measurements we can assume point-like morphology for the source. We performed the ETC calculations for three $H-$band magnitude values namely 20, 23 and 24\,mag. According to Figure~\ref{col-mag-diag} these magnitude values correspond to YSO masses of about 1.5, 0.40 and 0.20\,\Msun, respectively, for free-extinction 5\,Myr YSOs at 50\,kpc. 
In the Milky Way, these three values of magnitude correspond respectively to 0.2\,\Msun and brown dwarfs at 5\,kpc and A$\rm_V$=10 (typical of the Westerlund supermassive clusters); to 1, 0.13\,\Msun\, and brown dwarfs at 6\,kpc and A$\rm_V$=15 (typical of the RSGC supermassive clusters); and 1.7, 0.35, 0.17\,\Msun\, at 8.5\,kpc and A$\rm_V$=20 (typical of the supermassive clusters close the the Galactic center). 

\begin{figure*}[ht]
\centering
\includegraphics[width=1\textwidth ]{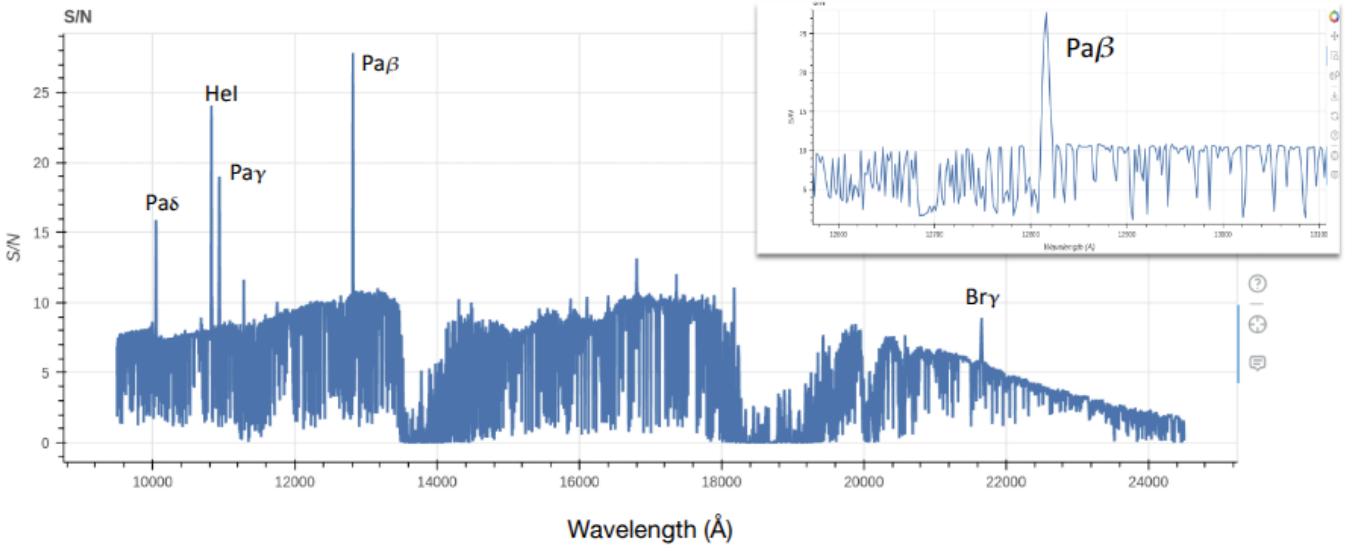}
\caption{Plot of S/N ratio versus wavelength as output from the ETC calculations for the
faintest YSO case ($H=$24\,mag) and the NEXUS calculation for a resolving power of 6000 
(see column 3 in Table~\ref{Table_S_N}). The other ETC parameters were fixed as explained
in Section~\ref{performance} and the adopted template spectrum is shown in Figure~\ref{template_Sz88A}.
Typical permitted emission lines are marked with the inset showing a region arounf the \pab\ line.}
\label{S_N_faint}
\end{figure*}

The results for the reference wavelength at 22000\,\AA\ are reported in Table~\ref{Table_S_N}, with the Nexus example for the faintest magnitude and resolution of 6000, shown graphically in Figure~\ref{S_N_faint}. In columns 5 to 8 of  Table~\ref{Table_S_N} we also report the limiting YSO mass corresponding to the three adopted $H$-band magnitudes; such mass was drawn from the ishochrones shown in Figure~\ref{col-mag-diag}.


\setlength{\tabcolsep}{3.5pt}
\begin{table*}[ht]
\centering 
\caption{Signal-to-noise ratio of the continuum according to the SHARP ETC at the reference wavelength 
of 22000\AA\, for the adopted template spectrum of the YSO Sz\,88A shown in Figure~\ref{template_Sz88A}. 
Estimates are provided for three $H-$band magnitude values (first column) and for both NEXUS-MOS 
(at two resolution values, columns 2 and 3) and VESPER-mIFU (column 4). See the text 
for the description of the adopted ETC parameters. Columns 5 to 8 provide the YSO mass corresponding 
to the magnitudes of column 1, and for the ishochrones shown in Figure~\ref{col-mag-diag}. 
Similarly, Columns 10-11 provide YSO mass for \Av=10\,mag and d=5\,kpc (SSC-1),  \Av=17\,mag and d=6\,kpc (SSC-2), and  \Av=20\,mag and d=8.5\,kpc (SSC-3)}
\vspace{0.5cm}
\begin{tabular}{|c| c| c| c| c| c| c| c | c | c | c |}
\hline
  $H$    &\multicolumn{2}{c|}{Nexus} & Vesper    & \multicolumn{7}{c|}{\Mstar/\Msun}    \\
   mag & $R_{2000}$ & $R_{6000}$ & $R_{17000}$ & \multicolumn{2}{c|}{\Av$=0$\,mag}  & \multicolumn{2}{c|}{\Av$=5$\,mag} & SSC-1 & SSC-2 & SSC-3  \\   
         &      &     &     &   1\,Myr& 5\,Myr &  1\,Myr  & 5\,Myr & 5\,Myr & 5\,Myr & 5\,Myr    \\
\hline \hline
  20     &  160 & 100 & 155 & 1.40    & 1.50 & 2.30 & 4.00  & 0.20 & 1.00 & 1.70 \\
  23     &  18  & 10  & 15  & 0.12    & 0.40 & 0.20 & 0.70  & $<0.1$ & 0.13  & 0.35 \\
  24     &  8   & 4   & 7   & $<$0.10 & 0.20 & 0.12 & 0.40  & $<0.1$ & $<0.1$  & 0.17\\
\hline
\end{tabular}
\label{Table_S_N}
\end{table*}

\begin{figure*}
    \centering
    \includegraphics[width=1\linewidth]{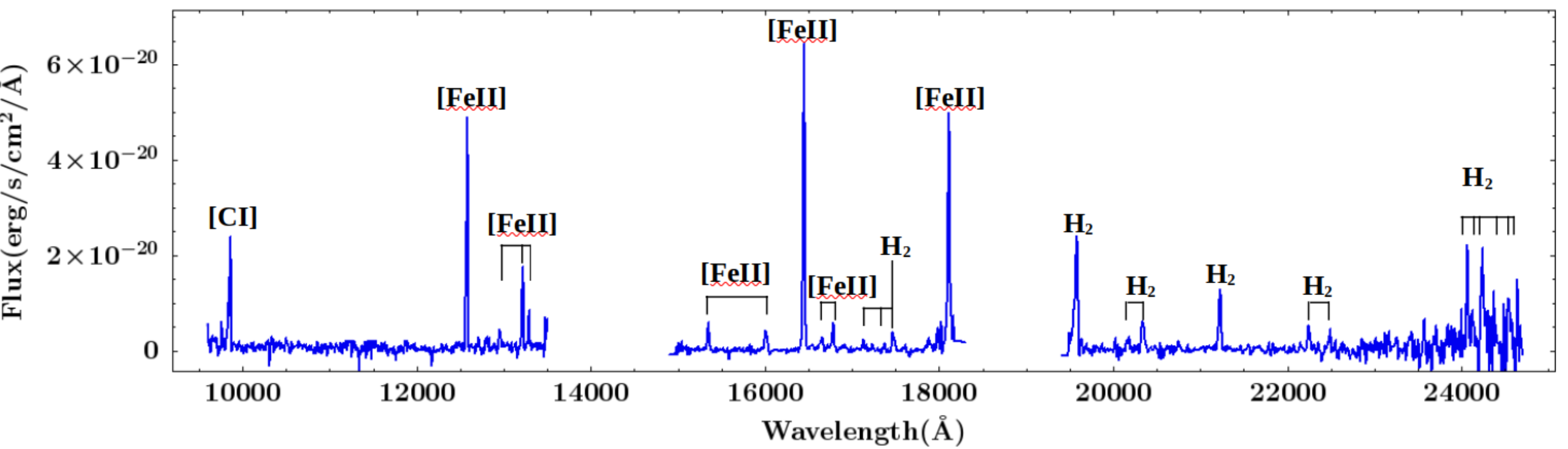}
     \caption{SOFI-ESO-NTT NIR spectrum of the HH\,111 jet in the Orion B molecular cloud (adapted from \citealt{nisini02}) shifted to the distance of the MCs. Note that the flux level of the [\ion{Fe}{ii}] and [\ion{C}{i}] lines should be reduced by a factor corresponding to the low metallicity, see text in subsection~\ref{etc_results} for details.  The main emission lines have been labelled.}
    \label{HH111inMC}
\end{figure*}

\subsubsection{Results from SHARP ETC}
\label{etc_results}

From the above ETC exercise we can assess that a reasonable limiting magnitude, yielding a S/N $>$ 10 in the spectral continuum, for our science case is H$\approx$24\,mag. According to the 1\,Myr isochrones shown in Figure~\ref{col-mag-diag}  this means that SHARP will be able to provide very good quality spectra of low-Z YSOs with masses down to $\approx$0.15\,\Msun\ located at a distance of  $\sim$50\,kpc, even for moderately extincted objects. This mass limit slightly 
rises to about 0.35\,\Msun\ for YSOs as old as 5\,Myr. For Milky Way SFRs like Sh2--284, about 10 times closer than the MCs (see Section~\ref{previous}), SHARP will provide high quality spectra of objects possibly down to the brown dwarf regime. 

In conclusion, SHARP observations shall allow us to acquire NIR spectra of YSOs with masses down to about 0.2\,\Msun\ with a S/N$\approx$15-20 (at $\lambda \approx 16000$\AA) in the continuum, with a resolving power higher than X-Shooter spectra, in dense regions of the outer MW Galaxy and in the MCs. Emission lines will have higher than the continuum S/N values (see Figure~\ref{S_N_faint}) allowing line flux measurements with a precision similar, but higher for the bright objects, to that of X-Shooter measurements of YSOs in nearby (d$<$500\,pc) SFRs. All this will allow one to characterize the star-disk interaction of low-mass YSOs in low-metallicity environments, in SFRs of the outer galaxy and in the MCs (see Section~\ref{proposal_goals}).

For the case of outflow/jets in the MCs, we used as templates spectra from the nearby ($\sim$400\,pc) HH\,111 flow in Orion, driven by a Class\,I protostar. However, we note that the expected line fluxes might be even larger for a similar flow located in the MCs, as the low metallicity might favour higher accretion and ejection rates.
We first shift the observed SOFI-ESO-NTT NIR spectrum (reported in \citealt{nisini02}) to the distance of the MCs (see Figure~\ref{HH111inMC}) and employ the ETC to compute the S/N of the different features with a 3600\,s on-source exposure. Note that the metallicity of the MCs is on the order of 0.5\,Z$_\odot$ and 0.2\,Z$_\odot$ for the LMC and SMC, respectively. Thus, in order to take into account the effects of low-metallicity, we adopt a value of 0.3 to scale the relevant atomic line fluxes in our calculations. This will impact the [\ion{Fe}{ii}] and [\ion{C}{i}] lines.

\begin{figure*}
    \centering
    \includegraphics[width=1\linewidth]{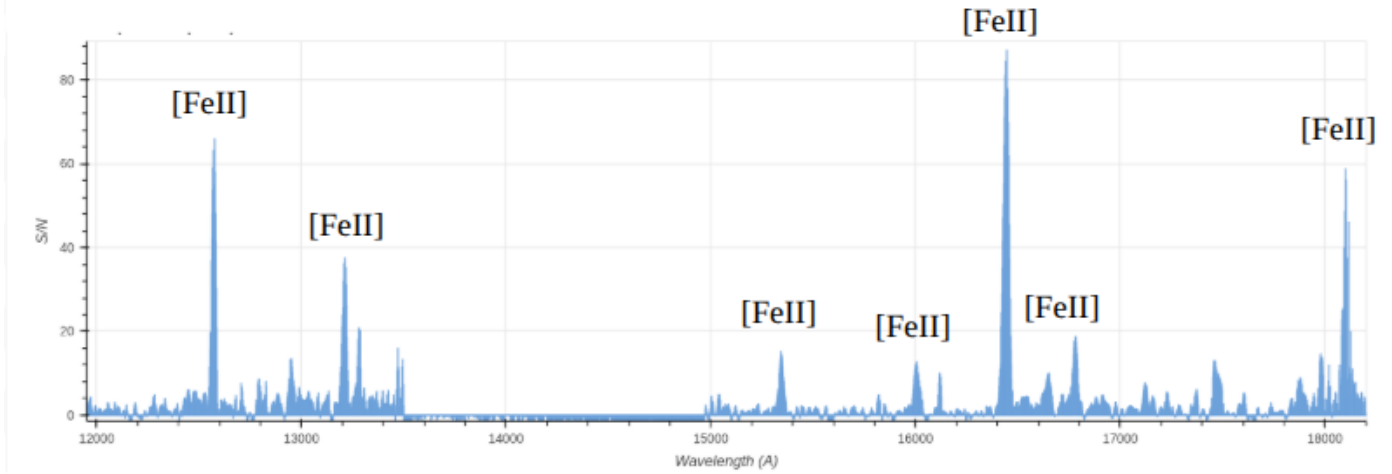}
     \caption{Plot of S/N ratio versus wavelength as output from the ETC calculations for the outflow in HH\,111 (H =24 mag) and the NEXUS calculation for a resolving power of 6000. The plot is restricted to the range (12000--18200\AA) of the [\ion{Fe}{ii}] lines of the template spectrum shown in Figure~\ref{HH111inMC}.  The latter was scaled to a factor 0.3 to consider the sub-solar metallicity (see subsection~\ref{etc_results}). }
    \label{SN_Fe}
\end{figure*}

The results, assuming similar ETC parameters as above, are as follows: the brightest [\ion{Fe}{ii}] line at 1.64\,$\mu$m (low-Z scaled flux $\sim$5.4$\times$10$^{-19}$ erg\,s$^{-1}$\,cm$^{-2}$) has a S/N of $\sim$\,85, while the weakest [\ion{Fe}{ii}] lines are at S/N $>$ 10 (see Figure~\ref{SN_Fe}). The brightest H$_2$ transition at 2.12\,$\mu$m (1-0\,S(1); flux $\sim$4.6$\times$10$^{-19}$\,erg\,s$^{-1}$\,cm$^{-2}$) has a S/N of $\sim$\,70. For the faintest H$_2$ lines (2-1\,S(2); flux $\sim$4$\times$10$^{-20}$ \,erg\,s$^{-1}$\,cm$^{-2}$)  S/N $\sim$\,5.
We stress that the average extinction towards the HH\,111 outflow is A$_V$=10\,mag. In principle, the brightest lines would be well detected (S/N$\sim$5) with extinction as high as A$_V$$\sim$40\,mag.


\subsection{Goals}
\label{proposal_goals}
Main goal of this science case is the investigation of the star-disk interaction process in large samples of 
YSOs in distant clusters and SFRs at low-Z environments, and compare the results with those of nearby solar 
metallicity and low-mass regions. Detailed goals are:

\begin{itemize}

\item investigate and characterize the low-mass (range 0.2 -- 2\Msun) YSOs;

\item characterize the accretion properties of the YSOs,
naturally allowing us to examine the accretion disk fraction in low-metallicity SFRs and supermassive clusters in comparison with that in solar metallicity SFRs and low-mass regions in the MW Galaxy;

\item investigate the relationships between the stellar physical parameters and the accretion properties
in low-Z and UV-dominated environments, such as the \Lstar\ vs. \Lacc\ and \Mstar\ vs. \Macc\ relationships, and compare the 
results with those of nearby solar metallicity and low-mass star forming regions. This will allow us to define whether disk dispersal 
in low-Z environments is more/less efficient than at solar metallicity and/or whether photo-evaporation
has a more important role and at which evolutionary stage;

\item detect atomic and molecular lines  (see Section~\ref{etc_results} and Figure~\ref{SN_Fe}) to investigate the physical and kinematical properties of jets and outflows in low-metallicity star forming regions.
This analysis can be applied to a large sample of sources having different stellar and accretion properties. In particular, the SHARP spectral resolution will enable us to distinguish and characterize the different kinematical components of the outflowing material, namely the atomic collimated jet and the wider angle molecular wind, as well as to investigate the relative importance of these components with age.

\item investigate the efficiency of the ejection and accretion processes at low-Z through the \Mout/\Macc\, parameter, which critically depends on the jet launching mechanism.
We will investigate whether H$_2$ molecular winds are driven by a  magneto-centrifugal mechanism (showing velocities from tens to hundreds km\,s$^{-1}$) or by  UV photoevaporation (with velocities of a few km\,s$^{-1}$) in low-metallicity and UV-dominated environments.

In the cases where position angle of the jets will be known, for instance from ground-based AO direct imaging
or from JWST images, it will be possible to align the NEXUS MOS slits along the jet direction. For cases where the
jet position angles are unknown Vesper-mIFU will provide information on the jets orientation and morphology.

\item accurate characterization of YSOs in young low-Z clusters and supermassive SFRS will allow us to define a sample of benchmark stars to be used as training datasets for machine learning supervised algorithms aimed to derive fundamental stellar parameters of YSOs from future photometric surveys such as Vera Rubin Legacy Survey of Space and Time (LSST) and Roman Telescope, as done for stars closer than  $\sim$1\,kpc in Tarantino et al. 2025 (in press). Combining Vera Rubin LSST photometric information, it will be possible to characterize the physical properties at work in YSOs related to accretion and ejection processes at different time scales (e.\,g. \citealt{bonito23, prisinzano23}).

\end{itemize}

\section{Conclusion}
\label{conclusion}

In this contribution we presented the SHARP scientific case on accretion and outflows in young stellar objects 
in distant low-metallicity and supermassive star forming regions.
Extreme ground-based adaptive optics, with a spatial resolution a factor 3 better than JWST, as well as sensitive multiplexing-IFU capabilities, uniquely offered by SHARP@ELT, are required to efficiently survey the whole area in low-Z and supermassive SFRs in 
the MW Galaxy  and in the MCs. 

With SHARP it will be possible to accurately characterize the young stellar populations and the star-disk interaction in YSOs with masses 
down to about 0.2\,\Msun\ in SFRs as distant as the Magellanic Clouds, and possibly down to the substellar regime in SFRs of the outer Milky Way Galaxy  ($<$10\,kpc scale). This represents a step forward in the investigation of the star-disk interaction process in low-metallicity YSOs and highly irradiated disks. It will also be possible to investigate differences between solar and sub- solar metallicity IMFs,
which may arise in low-Z regions of the outer Milky Way Galaxy and in the Magellanic Clouds (\citealt{rochau07,zinnkann24}).

In conclusion, SHARP observations of YSOs in distant low-metallicity and UV-dominated SFRs will lead to a step-change in the study of accretion and outflows 
in low-mass young stars hence, on protoplanetary disk evolution, as the spectrograph will enable investigation of targets that are several 
magnitudes fainter than those reachable with current instrumentation.

\vspace{1cm}

\small{
\noindent
{\bf Acknowledgements}

\noindent
We thank the referee for their very useful comments that helped to improve the paper.
We thank A. Armeni for discussions on photospheric tracers in the NIR. This work has been financially supported by Large Grant INAF-2024 “Spectral Key features of Young stellar objects: Wind-Accretion LinKs Explored in the infraRed (SKYWALKER)” , by
PRIN-MUR 2022 20228JPA3A ``The path to star and planet formation in the JWST era (PATH)" funded by NextGeneration EU and by INAF-GoG 2022 “NIR-dark Accretion Outbursts in Massive Young stellar objects (NAOMY)”. JMA and ACG acknowledge support from the INAF Mini-Grant 2023 ``Investigating the planet formation: initial conditions through the mass accretion rate on protostars". R.B. acknowledges support from the INAF Mini-Grant “Physical properties of Accreting young stellar objects: exploration of their light Curves and Emission (PACE)”.
LP acknowledges support from the Italian Ministero dell'Università e della Ricerca and
the European Union - Next Generation EU through project PRIN 2022 PM4JLH ``Know your little neighbours: characterising low-mass stars and planets in the Solar neighbourhood''.

}




{}

\end{document}